%% file: EMSE-Main.tex
\RequirePackage{fix-cm}

\documentclass[smallextended]{svjour3}

\usepackage{natbib}
\usepackage{balance}
\usepackage{fancyhdr}
\usepackage{amssymb}
\usepackage{amsmath}
\usepackage[figuresright]{rotating}
\usepackage{wrapfig}
\usepackage{multirow}
\usepackage{graphicx}
\usepackage{algorithm}
\usepackage{algorithmic}
\usepackage{tcolorbox}
\usepackage{url}
\usepackage{booktabs}
\usepackage{color}
\usepackage{moreverb}
\usepackage{balance}
\usepackage{amssymb,amsmath}
\usepackage{wrapfig}
\usepackage{multirow}
\usepackage{graphicx}
\usepackage{epstopdf}
\usepackage{tabularx}
\usepackage{bigstrut}
\usepackage{url}
\usepackage{ulem}
\usepackage{enumitem}
\usepackage{miniplot}
\usepackage{ifthen}

\usepackage{pifont}
\usepackage{bbding}
\usepackage{booktabs}
\usepackage{threeparttable}
\usepackage{array}
\usepackage{ragged2e}
\usepackage{etoolbox}
\usepackage{color}
\usepackage{colortbl}
\usepackage{cite}
\usepackage{amsmath,amssymb,amsfonts}
\usepackage{textcomp}
\usepackage{xcolor}
\usepackage{subfigure}
\usepackage{makecell}
\usepackage{booktabs} 
\usepackage{array} 
\usepackage{wrapfig}
\usepackage{parskip}

\makeatletter

\patchcmd{\NAT@citex}
  {\@citea\NAT@hyper@{%
     \NAT@nmfmt{\NAT@nm}%
     \hyper@natlinkbreak{\NAT@aysep\NAT@spacechar}{\@citeb\@extra@b@citeb}%
     \NAT@date}}
  {\@citea\NAT@nmfmt{\NAT@nm}%
   \NAT@aysep\NAT@spacechar\NAT@hyper@{\NAT@date}}{}{}

\patchcmd{\NAT@citex}
  {\@citea\NAT@hyper@{%
     \NAT@nmfmt{\NAT@nm}%
     \hyper@natlinkbreak{\NAT@spacechar\NAT@@open\if*#1*\else#1\NAT@spacechar\fi}%
       {\@citeb\@extra@b@citeb}%
     \NAT@date}}
  {\@citea\NAT@nmfmt{\NAT@nm}%
   \NAT@spacechar\NAT@@open\if*#1*\else#1\NAT@spacechar\fi\NAT@hyper@{\NAT@date}}
  {}{}

\makeatother

\newboolean{showcomments}
\setboolean{showcomments}{true}
\ifthenelse{\boolean{showcomments}}
{ \newcommand{\mynote}[2]{
    \fbox{\bfseries\sffamily\scriptsize#1}
    {\small$\blacktriangleright$\textsf{\emph{#2}}$\blacktriangleleft$}}}
{ \newcommand{\mynote}[2]{}}

\ifthenelse{\boolean{showcomments}}
{
  
}{
  \newcommand{\mynote}[2]{}
}

\begin{document}

\title{Towards an Understanding of Large Language Models in Software Engineering Tasks}

  \author{Zibin Zheng  \and Kaiwen Ning \and Qingyuan Zhong \and Jiachi Chen \thanks{Jiachi Chen is the corresponding author.}
  	 \and
	Wenqing Chen \and Lianghong Guo \and Weicheng Wang\and Yanlin Wang}

\institute{Zibin Zheng \at
	School of Software Engineering, Sun Yat-sen University, China\\
        Zhuhai Key Laboratory of Trusted Large Language Models\\
	\email{zhzibin@mail.sysu.edu.cn}
	\and
	Kaiwen Ning \at
	School of Software Engineering, Sun Yat-sen University, China \\
        Pengcheng Laboratory, China \\
        Zhuhai Key Laboratory of Trusted Large Language Models\\
	\email{ningkw@mail2.sysu.edu.cn}
	\and
 	Qingyuan Zhong \at
	School of Software Engineering, Sun Yat-sen University, China \\
        Zhuhai Key Laboratory of Trusted Large Language Models\\
	\email{zhongqy39@mail2.sysu.edu.cn}
	\and
	Jiachi Chen \at
	School of Software Engineering, Sun Yat-sen University, China\\
        Zhuhai Key Laboratory of Trusted Large Language Models\\
	\email{chenjch86@mail.sysu.edu.cn}
        \and
	Wenqing Chen \at
	School of Software Engineering, Sun Yat-sen University, China \\
        Zhuhai Key Laboratory of Trusted Large Language Models\\
	\email{chenwq95@mail.sysu.edu.cn}
        \and
	Lianghong Guo \at
	School of Software Engineering, Sun Yat-sen University, China \\
        Zhuhai Key Laboratory of Trusted Large Language Models\\
	\email{guolh8@mail2.sysu.edu.cn}
	\and
	Weicheng Wang \at
	School of Software Engineering, Sun Yat-sen University, China\\
        Zhuhai Key Laboratory of Trusted Large Language Models\\
	\email{wangwch7@mail2.sysu.edu.cn}
         \and
	Yanlin Wang \at
	School of Software Engineering, Sun Yat-sen University, China \\
        Zhuhai Key Laboratory of Trusted Large Language Models\\
	\email{wangylin36@mail.sysu.edu.cn}}

\date{Received: date / Accepted: date}
\maketitle
\thispagestyle{fancy}
\lhead{}
\chead{}
\rhead{}
\lfoot{}
\cfoot{}
\cfoot{\thepage}
\renewcommand{\headrulewidth}{0pt}
\renewcommand{\footrulewidth}{0pt}
\pagestyle{fancy}
\cfoot{\thepage}

  \input{abstract}


\input{Introduction1}
\input{Background}

\input{Methodology}
\input{RQ1}

\input{RQ2}

\input{RelatedWork}
\input{Conclusion}

\balance
\bibliographystyle{spbasic}
\bibliography{test}


\end{document}

%% file: abstract.tex
\begin{abstract}

Large Language Models (LLMs) have drawn widespread attention and research due to their astounding performance in text generation and reasoning tasks. Derivative products, like ChatGPT, have been extensively deployed and highly sought after. Meanwhile, the evaluation and optimization of LLMs in software engineering tasks, such as code generation, have become a research focus. However, there is still a lack of systematic research on applying and evaluating LLMs in software engineering. Therefore, this paper comprehensively investigate and collate the research and products combining LLMs with software engineering, aiming to answer two questions: (1) What are the current integrations of LLMs with software engineering? (2) Can LLMs effectively handle software engineering tasks? To find the answers, we have collected related literature as extensively as possible from seven mainstream databases and selected 123 timely papers published starting from 2022 for analysis. We have categorized these papers in detail and reviewed the current research status of LLMs from the perspective of seven major software engineering tasks, hoping this will help researchers better grasp the research trends and address the issues when applying LLMs. Meanwhile, we have also organized and presented papers with evaluation content to reveal the performance and effectiveness of LLMs in various software engineering tasks, guiding researchers and developers to optimize.

\end{abstract}

%% file: Introduction1.tex
\section{Introduction}
Large language models(LLMs) refer to neural network language models trained on massive text data, with model sizes reaching hundreds of billions or more parameters~\citep{G1_zhao2023survey}. The most advanced LLMs to date, such as GPT-4~\citep{N108, N139} and LaMDA~\citep{W25}, have demonstrated remarkable language comprehension and generation capabilities, able to perform well on a variety of natural language processing tasks, such as text summarization~\citep{N146}. The derivative products and applications of LLMs, such as ChatGPT~\citep{N96} and Claude~\citep{N156}, have also gained widespread attention in both academic and industrial communities. Given the outstanding performance of LLMs, there is a growing focus on exploring their potential in software engineering tasks, seeking new opportunities to address challenges in the field of software engineering~\citep{N85}.

The primary objective of software engineering is to develop high-quality and easily maintainable software products~\citep{N143}. This process involves multiple stages, including requirement analysis, design, development, testing, and maintenance~\citep{N144}. Throughout this process, engineers need to handle various software engineering tasks, such as understanding complex requirements, writing correct and reliable code, constructing comprehensive test cases, etc. In general, these software engineering tasks often rely on the expertise and experience of software engineers. This dependence not only incurs significant human resource costs, but also increases the difficulty of software development and maintenance. Additionally, with the increasing complexity of user demands and the emergence of new types of applications, engineers also face additional challenges when dealing with software engineering tasks, such as collaborative development, enhancing software quality, and shortening development cycles~\citep{N145}.

Therefore, researching how to enhance the level of automation in software engineering tasks to achieve more efficient software production and maintenance is a critical and widely discussed topic. Currently, numerous efforts are devoted to developing relevant tools and algorithms to assist engineers in completing these software engineering tasks. For instance, automated code generation~\citep{N149}, automatic test case generation~\citep{N150}, and vulnerability detection~\citep{N151} are some of the areas of focus. However, these individual automation methods often face challenges when applied universally~\citep{N16}, and some existing automated solutions may introduce new issues or errors. For example, automatically generated code may contain potential vulnerabilities~\citep{N80}, automatically generated test cases often fail to achieve comprehensive coverage~\citep{N91}, etc.

Fortunately, LLMs have great potential to solve the above problems due to their excellent performance on complex tasks such as text generation. LLMs usually need to be trained on a large-scale corpus containing a code base with multiple segments~\citep{G19_wei2022emergent}. This allows it to better learn the code syntax and semantics, and has the potential to be better equipped for tasks such as code completion and code summarization~\citep{W47}. Currently, as more and more LLMs designed for software engineering tasks are deployed~\citep{W23,W20,W33,W19,W28,W36,W47}, many research works focused on the application of LLMs in the software engineering domain~\citep{N10, N34, N45, N120,N79,N96}. The ability of LLMs such as generate high-quality code and high-coverage test cases has become a hot topic in the software engineering domain~\citep{N85}. However, in the existing literature, adequate systematic reviews and surveys have been conducted on LLMs applications in areas such as education~\citep{N122}, but a systematic review of the application and evaluation of LLMs in the field of software engineering is still missing.

In this study, our goal is to conduct a comprehensive review of the intersecting theme of LLMs and software engineering. To facilitate this, we initially gathered as much relevant literature as possible from six major academic databases, namely ACM Digital Library\footnote{https://dl.acm.org/}, IEEE Xplore Digital Library\footnote{https://ieeexplore.ieee.org/Xplore/home.jsp}, dblp\footnote{https://dblp.uni-trier.de/}, Elsevier Science Direct\footnote{https://www.sciencedirect.com/}, Google Scholar\footnote{https://scholar.google.com}, and arXiv\footnote{https://arxiv.org/}. Through the method of card sorting~\citep{N155}, we have eliminated duplicate and irrelevant literature. 
We focuses on papers published after 2022 because the definition and scale of "Large" in LLMs have evolved, making earlier literature potentially outdated. According to the collected papers we define transformer based language models with a parameter count greater than or equal to 0.8 billion (0.8b) as Large Language Models (LLMs).
After the screening process, we have obtained 123 relevant and valid research papers. With this study, we aim to address the following two questions:

\noindent \textbf{RQ1: What are the current works focusing on combining LLMs and software engineering?}

To answer these questions, we categorized the 123 selected papers according to the software engineering tasks involved. Based on the specific content of the software engineering tasks, such as code generation and vulnerability detection, we divided them into seven categories. They are Code Generation, Code Summarization, Code translation, Vulnerability Detection, Code Evaluation, Code Management and Q\&A Interaction. For each category, we elaborate on their definitions and examples, which can help researchers continue to discover and solve potential issues when applying LLMs to software engineering tasks.

\noindent \textbf{RQ2: Can LLMs truly help better perform current software engineering tasks? }

While LLMs have demonstrated outstanding performance in text generation tasks, their performance in software engineering tasks like code generation requires further validation. To address this issue, we conducted a selection of literature containing evaluations related to LLMs. Considering that the selected LLMs and software engineering tasks in these works may vary, we also organized and compiled this information during the screening process. Our findings indicate that currently, LLMs excel in tasks that demand an understanding of syntax, such as code summarization and code repair. However, their performance tends to be less satisfactory in tasks that require comprehension of code semantics, such as code generation and vulnerability detection. Nevertheless, we also observed that LLMs continue to make strides with each version and model iteration, indicating they still possess the potential to achieve better performance in the future.

The main contributions of this paper are:
\begin{itemize}
   \item We systematically reviewed the state-of-the-art work in the intersection of software engineering and LLM, a highly discussed topic. We manually selected 123 relevant works from many articles in six databases and conducted detailed organization and categorization. (Open source, \url{https://github.com/KevinHeiwa/LLM-SE-Paper-List});
   \item We categorized these tasks manually into seven types based on different software engineering tasks. For each category, we provided application examples of LLM and delved into detailed explanations. This can assist researchers in better identifying and addressing potential challenges when applying LLM to software engineering tasks;
   \item We have comprehensively collected and compiled the performance of LLM in various software engineering tasks. We have provided an exposition and analysis of LLM's performance in these software engineering tasks, as well as the reasons for variations observed among different studies. This can assist developers in optimizing LLM more effectively.
\end{itemize}

The organization of this paper is as follows. In  Section~\ref{sec:background}, we provide background knowledge on LLMs; In Section~\ref{sec:LLM Meet SE}, we summarize and categorize the collected literature and propose an answer to research question 1 (RQ1); In Section~\ref{sec:LLM Better}, we address research question 2 (RQ2); In Section~\ref{sec:related work}, we compiled the performance of different LLMs across various benchmarks and literature to address research question 3 (RQ3); Finally, in Section~\ref{sec:conclusion}, we summarize the entire paper.

%% file: Background.tex
\section{Background}
\label{sec:background}
In this section, we will introduce the background of large language models, including the transformer model, the architecture of large language models, and their emergent capabilities.

\subsection{Transformer}
Currently, the mainstream large language models are based on the Transformer~\citep{G5_vaswani2017attention}  model, such as GPT-3~\citep{G2_brown2020language} and PaLM~\citep{G3_chowdhery2022palm}. Compared to traditional deep learning model structures like recurrent neural networks (RNNs) and convolutional neural networks (CNNs), the Transformer model relies on a attention mechanism, allowing for parallel computation and more effective capture of long-range dependencies. This provides a foundation for effectively training large language models on large-scale text data.

The Transformer model first introduced by Vaswani et al~\citep{G5_vaswani2017attention} in 2017, is a sequence-to-sequence model consisting of an encoder and a decoder. Both the encoder and decoder are composed of multiple identical blocks stacked together. Each block in the encoder primarily includes a multi-head self-attention module and a position-wise feed-forward network (FFN), with residual connections~\citep{G7_he2016deep} and layer normalization~\citep{G8_ba2016layer} applied to enable deeper model building.~\citep{G6_lin2022survey} In the decoder, cross-attention modules are inserted between the multi-head self-attention modules and the position-wise FFNs, allowing for the incorporation of information from the encoder. Notably, the self-attention modules in the decoder are modified to prevent attending to subsequent positions.

The core component of the Transformer is the attention mechanism, which allows the model to weigh the importance of different words or tokens in a sequence when generating or understanding the context. By attending to relevant parts of the input sequence, the Transformer model can effectively model the relationships between words and capture rich contextual information. Additionally, the attention mechanism does not involve sequential operations, enabling parallel computation and resulting in higher efficiency during training and inference in the Transformer model.

\subsection{Model Architecture}
The Transformer architecture, proposed by Vaswani et al. in 2017~\citep{G5_vaswani2017attention}, has emerged as the leading choice for developing large language models (LLMs) due to its exceptional parallelizability and capacity~\citep{G1_zhao2023survey}. This scalability allows language models to be expanded to include hundreds or even thousands of billions of parameters, enabling them to capture more complex language patterns and improve performance on various tasks. In general, large language models can be categorized into three main architecture types: encoder-decoder structures, causal-decoder, and prefix decoder~\citep{G1_zhao2023survey}, each with its own characteristics and applications.

\textbf{Encoder-decoder Architecture:} The vanilla Transformer proposed in~\citep{G5_vaswani2017attention} is based on encoder-decoder architecture, which comprises separate encoder and decoder components. The encoder processes the input sequence and captures its latent representation, which is then used by the decoder to generate the output sequence autoregressively. This architecture is well-suited for tasks involving sequence-to-sequence mapping, such as machine translation, text summarization, and dialogue generation. Encoder-decoder pretrained model. Encoder-decoder pretrained models, such as BART~\citep{G10_lewis2019bart} and T5~\citep{G9_raffel2020exploring}, have demonstrated excellent performance across various downstream tasks. However, with the development of LLM there are only a few large language models based on the encoder-decoder architecture, such as Flan-T5~\citep{G11_chung2022scaling} and CodeT5+~\citep{N16}.

\textbf{Causal Decoder Architecture:} The causal decoder architecture is commonly implemented as a stack of decoder layers. It utilizes a diagonal mask matrix, allowing each token to only have access to information from preceding tokens. This constraint ensures a unidirectional and autoregressive generation process. The GPT series model, initially introduced by OpenAI~\citep{G12_radford2018improving,G13_radford2019language,G2_brown2020language}, represents one of the most prominent examples of the causal decoder architecture. While GPT~\citep{G12_radford2018improving} and GPT-2~\citep{G12_radford2018improving} did not exhibit the same level of performance as GPT-3~\citep{G2_brown2020language}, with the increase in model size and the amount of data used for pretraining, GPT-3~\citep{G2_brown2020language} showcased a remarkable few-shot capability that earlier models did not possess. Today, the causal decoder architecture has become the prevailing choice for large language model architectures, giving rise to a wide range of powerful LLMs such as PaLM~\citep{G3_chowdhery2022palm}, LLaMA~\citep{G4_touvron2023llama}, OPT~\citep{G14_zhang2022opt}, Bloom~\citep{G15_scao2022bloom}. 
The causal decoder architecture and the prefix decoder architecture, which will be discussed next, are collectively referred to as decoder-only architecture~\citep{G1_zhao2023survey}.

\textbf{Prefix Decoder Architecture:} The prefix decoder, similar to the causal decoder architecture, consists of decoder layers. However, the key distinction is in their attention mechanism. The prefix decoder utilizes bidirectional attention for the prefix tokens, incorporating information from both preceding and succeeding tokens. In contrast, unidirectional attention is applied only to the generated tokens, ensuring a unidirectional flow of information during the generation process. This combination of attention mechanisms in the prefix decoder enables flexible and controlled generation, conditioned on both the prefix and the generated tokens. Some commonly known models based on the prefix decoder architecture include U-PaLM~\citep{G16_tay2022transcending} and GLM-130B~\citep{G17_zeng2022glm}.

\subsection{Emergent Abilities}
According to the scaling law of large language models~\citep{G18_kaplan2020scaling}, as the model parameters and training data increase, the model's capacity and capabilities also improve. When scaling surpasses a certain threshold, LLMs demonstrate emergent abilities that are not present in smaller models~\citep{G19_wei2022emergent}. These emergent abilities are considered the most notable characteristic that sets large models apart from their smaller counterparts. Such as in-context learning~\citep{G2_brown2020language}, instruction following~\citep{G20_sanh2021multitask,G21_ouyang2022training,G22_wei2021finetuned}, and step-by-step reasoning~\citep{G28_shanahan2022talking}.

%% file: Methodology.tex
\section{Methodology}
\label{sec:methodogy}
\begin{figure}
\includegraphics[width=0.9\textwidth]{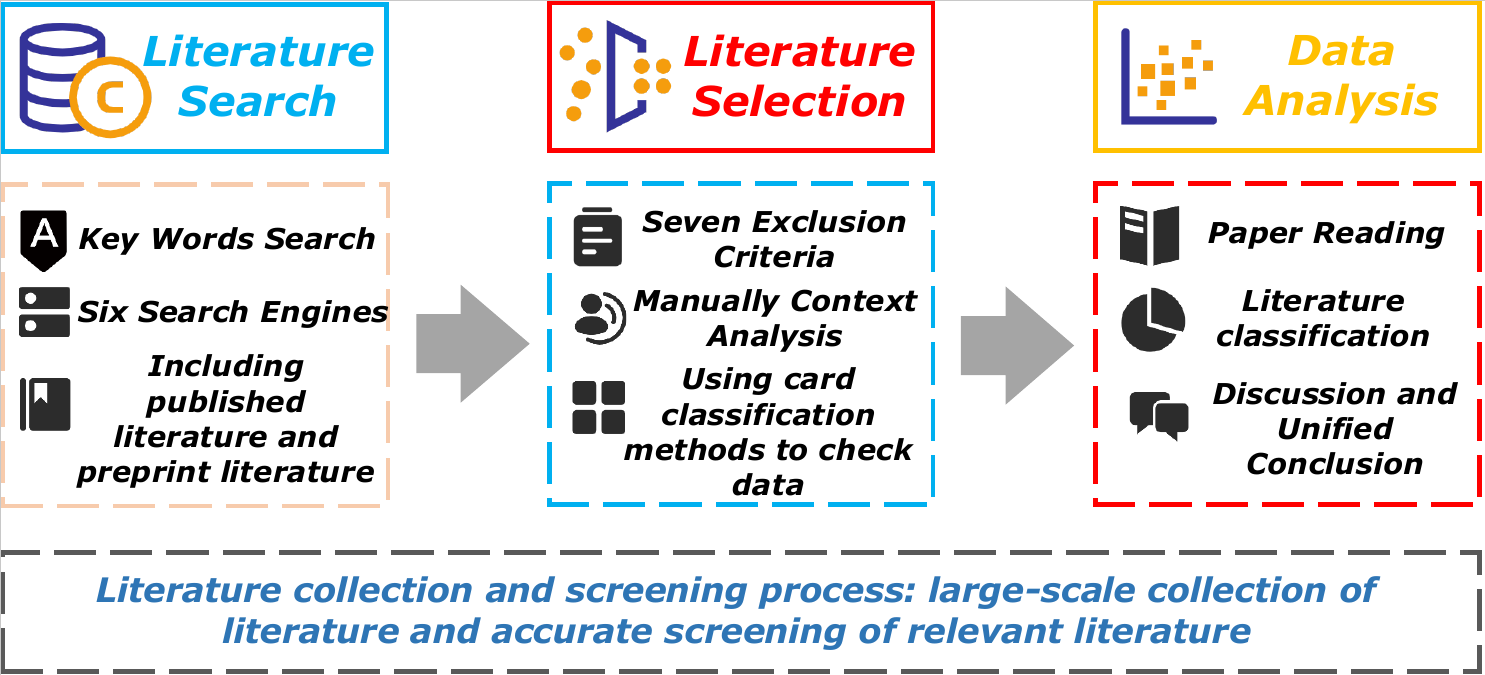}
\caption{Overview of methodology design.}
\label{fig:dataprocess}
\end{figure}

In this section, we introduce the detailed steps of conducting a literature review. We follow the method provided by~\citep{N141,Nn23} for the literature review. Generally speaking, it consists of three steps: literature search, literature screening, and data analysis. As shown in Fig.~\ref{fig:dataprocess}.

\subsection{Literature Search}
Based on the previous literature review~\citep{N142}, we have selected six search engines: ACM Digital Library, IEEE Xplore Digital Library, dblp, Elsevier Science Direct, Google Scholar, and arXiv. These search engines allow us to find peer-reviewed research papers published in journals, conferences, workshops, and symposiums. Additionally, they provide access to a considerable number of preprint papers and the latest industry developments.

We conducted searches using the following six keywords: ``SE LLM," ``Software Engineering Large Language Model," ``Software Engineering LLM," ``SE Large Language Model," ``Code LLM," and ``Code Large Language Model" on the aforementioned six paper databases. The obtained results are presented in Table~\ref{tab:paperdata}. It is worth noting that there might be a significant number of duplicate papers and irrelevant articles resulting from different keyword searches within the same engine or the same keyword across different engines. Therefore, we need to manually screen and select these papers, which is known as literature screening or literature selection.

\begin{table}[t]
\caption{Number of keyword searches returned by each search engine.}
\setlength{\tabcolsep}{2mm}
\renewcommand\arraystretch{1.3}
\label{tab:paperdata}
\begin{tabular}{p{40pt} p{20pt} p{40pt} p{40pt} p{40pt} p{40pt} p{40pt} }
\toprule
                            & \textbf{SE LLM} & \textbf{Software Engineering Large Language Model} & \textbf{Software Engineering LLM} & \textbf{SE Large Language Model} & \textbf{Code LLM} & \textbf{Code Large Language Model} \\ \midrule
ACM Digital Library         & 4983   & 55907                                     & 39032                    & 54532                   & 27894    & 54990                                          \\ \midrule
IEEE Xplore Digital Library & 0      & 240                                       & 7                        & 14                      & 14       & 328                                            \\ \midrule
dblp                        & 8      & 4                                         & 1                        & 105                     & 7        & 62                                             \\ \midrule
Elsevier Science Direct     & 56     & 11273                                     & 54                       & 7063                    & 64       & 20649                                          \\ \midrule
Google Scholar              & 10500  & 16400                                     & 4020                     & 25400                   & 11400    & 17700                                          \\ \midrule
arXiv                       & 5      & 182                                       & 70                       & 17                      & 463      & 1461                                           \\ \bottomrule
\end{tabular}
\end{table}

\subsection{Literature Selection}

During this stage, we need to eliminate not only duplicate papers but also irrelevant ones. For instance, some papers may primarily focus on LLM or the field of software engineering but do not mention LLM or ``Large Language Model" in their abstracts. Additionally, since the definition of ``Large" in LLM is subject to change, some earlier literature might have been considered LLM at the time but may not meet the criteria from perspective today~\citep{G1_zhao2023survey,G6_lin2022survey,N85,N100,N101,N102}. Therefore, we have excluded research conducted before 2022. We applied the following seven exclusion criteria to screen the literature:

\textbf{Exclusion Criteria}
\begin{itemize}
    \item Studies are not written in English.
    \item Master or Ph.D. theses.
    \item Keynote papers.
    \item Studies not related to LLM.
    \item Studies not related to software engineering.
    \item Duplicate literature.
    \item Studies up to 2022 (not including 2022).
\end{itemize}

To demonstrate the rationale behind excluding works before 2022, we collected and organized some of the important articles on language models (LMs) published in top software engineering conferences, namely ICSE, ISSTA, FSE, and ASE, between 2017 and 2022. The results are presented in Table~\ref{tab:paper_ICSE_PLM_1}, where we found a total of 19 relevant articles.

From Table~\ref{tab:paper_ICSE_PLM_1}, we can observe the following: Firstly, although the Transformer architecture was introduced in 2017, the application of Transformer-based pre-trained language models (PLMs) in SE tasks began around 2021. Secondly, previous articles primarily focused on "pre-trained language models" rather than LLMs. These two concepts have distinct differences. Lastly, LLMs started receiving attention and gradually being applied to software engineering tasks around 2022.

Furthermore, we can see that the research emphasis on PLMs and LLMs in SE tasks is quite different. For example, in Table~\ref{tab:paper_ICSE_1}, we can observe that code generation tasks are an important topic in LLM research but are rarely mentioned in studies conducted before 2022. This is due to the fact that the popular PLM architecture before 2022 was Encoder-Decoder, which significantly differs from the decoder-only architecture commonly used in current LLMs. This is the reason why we focus only on works after 2022.

\begin{table}[h]
\small
\caption{Articles about pre-trained language models between 2017 and 2022 in top software engineering conferences.}\label{tab:paper_ICSE_PLM_1}
\scalebox{1}{
\begin{tabular}{lllll}
\toprule & \textbf{ISSTA} & \textbf{ICSE} & \textbf{ASE} & \textbf{FSE} \\
\midrule 2017 & 0 & 0 & 0 & 0 \\
\midrule 2018 & 0 & 0 & 0 & 0 \\
\midrule 2019 & 0 & 0 & 1 & 0 \\
\midrule 2020 & 0 & 0 & 0 & 0 \\
\midrule 2021 & 0 & 2 & 2 & 0 \\
\midrule 2022 & 2 & 7 & 4 & 2 \\
\bottomrule
\end{tabular}
}
\end{table}

In this study, we are specifically focused on the intersection of LLM and software engineering. Therefore, we will exclude papers that solely focus on LLM or software engineering. We are interested in the following topics:

\textbf{Inclusion Topics}
\begin{itemize}
    \item LLM in Software Engineering.
    \item Application of LLM in Software Engineering (for example, code generation).
    \item Empirical Studies of LLM on Software Engineering Tasks
\end{itemize}

To improve the accuracy of the literature screening process, we will use the card sorting method to evaluate the collected data. Card sorting is a common technique used to assess data and derive categories from it. There are three types of card sorting methods: open card sorting, closed card sorting, and hybrid card sorting. Among these three types, closed card sorting involves predefined categories~\citep{N141}. In our case, we applied closed card sorting to select relevant papers since we have only two categories: relevant and irrelevant. Each card will have a title (paper title) and a description (paper abstract). By using this method, we can systematically evaluated and categorized the papers based on their relevance to our research.

Six experienced researchers, including one non-coauthors, independently conducted a thorough review of the search engine results from the six databases. After individually organizing the papers, they engaged in a collaborative discussion to align their findings. Through this rigorous process, we ultimately identified 123 relevant papers that met the criteria for inclusion in our study.

\subsection{Data Analysis}
The definition of ``large" in LLM changes over time. For this reason, we have filtered out work that does not meet the definition of LLM in~\citep{G1_zhao2023survey} and ensured that all such work will be made public after 2022. We used open card sorting to help find the answers to these two RQs. We read the article carefully and looked closely for answers to the same two questions shown in Table~\ref{tab:RQdata}, i.e.,(1). What are the current works focusing on combining LLMs and software engineering? (2). Can LLMs truly help to better execute current software engineering tasks? If we cannot find any answers from a paper, then that paper will be removed from our list.

\begin{figure}
\includegraphics[width=0.75\textwidth]{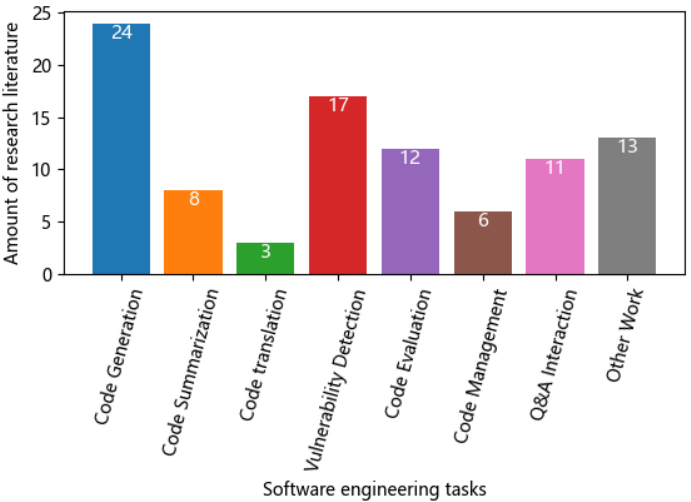}
\caption{Number of literature on different software engineering tasks.}
\label{fig:paperdata}
\end{figure}

For the answers to (1), we primarily examined whether the papers mentioned the application of LLM in software engineering tasks. We organized this information and categorized the literature based on the types of tasks. The number of papers included for each task is shown in Fig.~\ref{fig:paperdata}. Our selection of software engineering task classification was based on the main content of the surveyed articles. For a specific software engineering task, if there were only two articles related to that task, we categorized it separately. Otherwise, we grouped it under ``Other Work."

To validate our classification, we organized the research papers from ICSE 2024~\footnote{https://conf.researchr.org/track/icse-2024/icse-2024-research-track?\#event-overview}, a highly recognized conference in the field of software engineering, as shown in Table~\ref{tab:paper_ICSE_1}. We can observe that there were a total of 40 papers related to LLM in ICSE 2024, which can be categorized into the seven classes mentioned in the paper. This indicates that the software engineering tasks we selected are currently the most prominent in LLM and SE research and encompass a significant portion of software engineering tasks. Furthermore, concerning benchmarking, datasets, and similar work, we do not categorize them as software engineering tasks. Instead, we classify them based on their research targets. For example, a benchmark for code generation would still fall under the code generation task. The definitions of different tasks are provided in Table~\ref{tab:RQ1}.

\begin{table}[t]
\small
\caption{Papers related to LLM in ICSE 2024.}\label{tab:paper_ICSE_1}
\scalebox{1}{
\begin{tabular}{lll}
\toprule \textbf{Tasks} & \textbf{Numbers of Paper} & \textbf{Note} \\
\midrule Code Generation & 13 & Includes code improvements \\
\midrule Q\&A Interaction & 5 & $/$ \\
\midrule Code Translation & 1 & $/$\\
\midrule Code Summarization & 4 & Includes code comprehension \\
\midrule Code Evaluation & 3 & Includes code clone \\
\midrule Vulnerability Detection & 9 & Included Tests \\
\midrule Code Management & 4 & Includes logs \\
\midrule Other Works & 1 & $/$\\
\midrule Total & 40 & $/$ \\
\bottomrule
\end{tabular}
}
\end{table}

We can see that this is a relatively large number of Code Generation-oriented studies with 24 papers; conversely, Code translation-oriented is the least, with 3 papers. The number of papers in remaining 6 software engineering tasks are similar, with the least being Code Managenment-oriented work, with 6 papers. And Vulnerability Detection oriented tasks are the most with 17 papers. The specifics of these tasks are presented in \ref{sec:LLM Meet SE}.

For the answers to (2), we focused on reading whether the papers provided critical or explicit opinions on the performance of LLM in software engineering. Particularly, we examined instance studies and survey articles in this regard.

\begin{table}[t]
\caption{Data Collection for Each RQ.}
\setlength{\tabcolsep}{2mm}
\renewcommand\arraystretch{1.5}
\label{tab:RQdata}
\begin{tabular}{p{40pt} p{200pt}}
\toprule
\textbf{RQs} & \textbf{Type of Data We Collected}                                                                                                                                 \\ \midrule
RQ1 & What are the current works focusing on combining LLMs and software engineering? Such as code summarization, code translation, code generation, interaction with software developers, etc. \\  \midrule
RQ2 & Can LLMs truly help to better execute current software engineering tasks? Such as performance status, defects, potential, completion, deficiencies, etc.                   \\ \bottomrule
\end{tabular}
\end{table}

\begin{table*}[t]
\caption{The Definition of Seven Types of Software Engineering Tasks and the Role of LLMs.}
\setlength{\tabcolsep}{2mm}
\renewcommand\arraystretch{1.5}
\label{tab:RQ1}
\begin{tabular}{p{40pt}  p{140pt}   p{140pt}}
\toprule
\textbf{Task}  & \textbf{Definition}  & \textbf{The possible role of LLMs}  \\ 
\midrule
Code Generation  & Automatically Generating source code based on user requirements and specified constraints. Includes code completion, code enhancement, and other related tasks.  & (auxiliary) Generate code or provide developers with ideas and programming 'starting points',etc.   \\ \midrule

Code Summarization      &To automatically generate clear, accurate, and useful code comments to aid developers in understanding and maintaining code. Includes tasks such as code comprehension.& Code summaries that assist with different granularity (such as functions) or explain the intent of the code,etc.  \\ \midrule
Code translation        & Converting code between different programming languages without altering its functionality or logic.          & Auxiliary code conversion, Reverse engineering,etc.                                                            \\ \midrule
Vulnerability Detection & To identify and fix code errors that may cause program crashes, performance degradation, or security issues. Includes tasks such as vulnerability mining, vulnerability remediation and code testing.              & Check for potential vulnerabilities in the code, etc.                                                            \\ \midrule
Code Evaluation         & To perform static analysis on the code to identify potential issues and improvement points. Includes tasks such as Code Clone, Code Smell, etc.                                   & Generate test cases or test code performance, usability, and other indicators, etc.                               \\ \midrule
Code Management         & Manage information such as code versions and developers. Contains log-related tasks                                                                     & Team collaborative development, version control, etc                                                       \\ \midrule
Q\&A Interaction   & Interaction between software developers and LLM                                                                              & Program assistant, Prompt engineering, etc                                                                  \\ 
\midrule
Other Works   & Some other work, such as researching copyright and ethical issues in LLM-generated code.                                                                             & Unknown.          \\              \bottomrule
\end{tabular}
\end{table*}

%% file: RQ1.tex
\section{LLMs in Software Engineering Tasks}
\label{sec:LLM Meet SE}

In this section, we primarily respond to \textbf{RQ1}. We categorize the collected work into seven categories based on different tasks, and then provide separate explanations for each category, as shown in Table~\ref{tab:RQ1}.

In this paper, we consider models with a parameter count greater than or equal to 0.8 billion (0.8b) as Large Language Models (LLMs). We arrived at this definition based on our investigation of the majority of open-source LLM papers from ICSE 2024, as shown in Table~\ref{tab:paper_ICSE_LLM}. We can see that, except for OpenAI's models, which are not open-source, we cannot determine their parameter counts. Among the open-source models, the one with the smallest parameter count is Flan-T5-small, with 80 million (80M) parameters. Therefore, we have defined LLM based on this model as the standard. Among the peer-reviewed articles we investigated, the smallest parameter count used for LLM was 80M (\citep{N1n1}, Flan-T5-small).

\begin{table}[t]
\small
\caption{LLM with minimum number of parameters of articles in ICSE 2024.}\label{tab:paper_ICSE_LLM}
\scalebox{1}{
\begin{tabular}{ll}
\toprule \textbf{Paper} & \textbf{LLM with minimum number of parameters in the article}  \\
\midrule \citep{Nnnn1} & Flan-T5-small-80M \\
\midrule \citep{Nnnn9} & text-embedding-babbage-001 \\
\midrule \citep{Nnnn2} & GPT-3.5 \\
\midrule \citep{Nnnn10} & code-davinci-002 \\
\midrule \citep{Nnnn3} & codet5-base-232M \\
\midrule \citep{Nnnn11} & CodeParrot-small-110M \\
\midrule \citep{Nnnn4} & UnixCoder \\
\midrule \citep{Nnnn12} & ChatGPT \\
\midrule \citep{Nnnn5} & ChatGPT \\
\midrule \citep{Nnnn13} & GPT-4 \\
\midrule \citep{Nnnn6} & CodeGen-Multi-16B \\
\midrule \citep{Nnnn14} & PanGu-Coder-300M \\
\midrule \citep{Nnnn7} & ChatGPT \\
\midrule \citep{Nnnn15} & ChatGPT \\
\midrule \citep{Nnnn8} & GPT-3 \\
\midrule \citep{Nnnn16} & code-davinci-002 \\
\bottomrule
\end{tabular}
}
\end{table}

\subsection{Code Generation}

\noindent \textbf{Definition:} Code generation, also known as program synthesis, is the process of automatically generating source code based on user requirements and specified constraints~\citep{N53}. In most cases, it involves transforming text into code. When applying LLM for code generation, LLM will automatically generate code based on the requirements provided by the user. It should be noted that code completion, code refactoring, and code augmentation also fall under the category of code generation tasks.

Traditional code generation typically involves two steps: first, parsing the user's requirements and constraints and converting them into an intermediate representation, such as an abstract syntax tree (AST)~\citep{N157} or intermediate representation (IR)~\citep{N158}; then, generating the target code based on this intermediate representation. The first step is requirement understanding, which involves transforming natural language into a formal, computer-understandable representation; the second step is code synthesis, which involves generating code based on this representation~\citep{N24}.

Code generation heavily relies on search and inference, systematically searching and reasoning to find code that satisfies the given requirements within the entire code space~\citep{W33}. LLMs has demonstrated impressive capabilities in text generation tasks, attracting significant research efforts to evaluate and improve the performance of LLM in code generation tasks~\citep{N37}.

These research efforts can be roughly categorized into two main themes. The first theme mainly evaluates or discusses the capabilities of LLMs in code generation tasks or specific contexts of code generation~\citep{N4,N74,N173,N174}. The evaluation perspectives vary, with some focusing on the correctness of code generation in different programming languages~\citep{N95,N80,N113}, while others propose new benchmark frameworks or testing methods to better evaluate the code generated by LLMs~\citep{N42, N76}, providing directions for improving LLMs in this task.

However, it is important to note that no current technology, including LLMs, can guarantee that automatically generated code is always completely usable, as there may be obvious or subtle errors in the code. Therefore, the second theme of these research efforts is to enhance the code generation capabilities of LLMs. This includes automatically fixing bugs in code generated by LLMs~\citep{N111,N89,N23}, improving the quality of code generated by LLMs~\citep{N41,N61,N120,N117,N64,N49,N24}, addressing security and reliability concerns~\citep{N118,N78,N114}, enhancing the efficiency of LLM code generation~\citep{N15,N31,N112,N170,N171,N172}, watermarking techniques for code generation\citep{N32,N169}, and among others.

\subsection{Code Summarization}

\noindent \textbf{Definition:} Code summarization is an important technique for program understanding and automatic documentation generation. When using LLM for code summarization, the main goal of this task is to automatically generate clear, accurate, and useful code comments to aid developers in understanding and maintaining code. It should be noted that code summarization tasks encompass both code understanding and code summarization tasks.

Code summarization can be performed at different granularities, such as line-level, function-level, or module-level. Different levels of summarization require handling different contextual information. For example, function-level summarization may need to understand the inputs, outputs, functionality, and invocation relationships of functions, while module-level summarization may require understanding the functionality and interfaces with other modules.

Traditional code summarization methods mainly rely on rules and templates, using specific rules and templates to extract and represent key information from code. However, these methods often require a significant amount of manual effort and have poor adaptability to new programming languages and frameworks. With the advancement of machine learning, neural network-based code summarization methods have gained attention. These methods use neural networks to learn the complex mapping between code and text. Additionally, some methods employ encoder-decoder architectures, where the encoder is used to understand the source code and the decoder is used to generate natural language descriptions.

The use of Large Language Models (LLMs) is a recent research direction in code summarization. These models, such as OpenAI's GPT series, have also received considerable attention in the specific task of code summarization. Some works evaluate the code summarization capabilities of different LLMs, such as ChatGPT~\citep{N14}, Codex~\citep{N74,N122}, GPT-3~\citep{N75}, among others. Other works assess the capabilities of LLMs in code summarization from different perspectives, such as causal reasoning abilities~\citep{N28}, effectiveness with minimal training~\citep{N68}, and interpretability~\citep{N52}. Although these works provide incomplete conclusions, they generally recognize the strong potential of LLMs.

There are also works that utilize LLMs for code summarization to construct documentation, generate test cases, or address limitations in LLM-based code summarization~\citep{N124,N34}. For example, enhancing the robustness of LLM-based code summarization~\citep{N79} or improving the interaction capabilities with developers~\citep{N60}.

\subsection{Code Translation}

\noindent \textbf{Definition:} Code translation, also named code conversion, refers to the process of converting code between different programming languages without altering its functionality or logic.When using LLM for code transformation, LLM will take the given code and transformation requirements from the user and convert the code to the target language.

Traditional code translation methods often require substantial manual effort, implementing specific syntax and semantic rules through hard coding. Moreover, for new or less common programming languages, it may necessitate the redevelopment and maintenance of translation rules. Currently, the impressive natural language translation capabilities exhibited by LLMs have been recognized~\citep{N35}. However, there is limited focus on the performance of LLMs in code conversion tasks in current research. Pan et al.~\citep{N175} present a large-scale empirical study to investigate the ability of LLMs, including general and code LLMs, for code translation in five languages, They are C, C++, Go, Java, and Python.

Pearce et al.\citep{N70} studied the capability of LLMs in software reverse engineering. The study explored Codex's ability to identify the purpose, functionality, and important variable names or values in code, thus evaluating the decompilation ability of LLMs. Additionally, Lin et al.\citep{N77} proposed a Cross-language Code representation with a large-scale pre-training (XCode) method and further introduced a Shared Encoder-Decoder (SED) architecture.

Currently, there is relatively little research on LLMs in code translation tasks, and applying LLMs to code translation tasks still faces many challenges. Firstly, code correctness and precision are crucial, as even small translation errors can render the generated code non-functional. Secondly, acquiring a large amount of high-quality source code and target code pairs is challenging, which may limit the model's learning and generalization capabilities. Thirdly, further research is needed on evaluating the performance of code translation models, as in many cases, there can be multiple different implementations of the same functionality in code. These issues require further exploration and resolution in future research.

\subsection{Vulnerability Detection\& Repair}

\noindent \textbf{Definition:} Code vulnerability detection and repair is an important task in the field of software engineering, crucial for improving the reliability and security of software. The main goal of this task is to identify and fix code errors that may cause program crashes, performance degradation, or security issues. When using LLM for vulnerability detection, LLM analyzes the target code for vulnerabilities, provides vulnerability reports, or fixes the detected vulnerabilities. It is worth noting that within the scope of this article's definition, tasks such as vulnerability mining and test case generation also fall under this task.

Due to the significance of this task, various software and hardware-based detection methods have been developed. Traditional vulnerability detection methods mainly include static analysis and dynamic analysis. Static analysis analyzes the source code before program execution to identify potential errors and vulnerabilities, while dynamic analysis analyzes program behavior during runtime to identify actual errors and vulnerabilities. These methods have their own advantages and disadvantages: static analysis can cover all code paths but may produce a large number of false positives, while dynamic analysis can precisely pinpoint actual errors but may miss errors that are not triggered in the test cases.

Recently, LLMs have been used in code vulnerability detection and repair tasks due to their advanced semantic understanding capabilities, which are effective in addressing logical flaws and linguistic issues in natural language. Similarly, in this task, current research can be roughly categorized into three types. The first type evaluates the capabilities of different LLMs in this task~\citep{N20, N134, N71, N55, N176}, for example, Khoury et al.~\citep{N126} evaluated the security of ChatGPT in code generation tasks. The second type focuses on improving the correctness and performance of LLMs in vulnerability detection~\citep{N58, N137}, such as combining LLMs with formal verification techniques~\citep{N33}, incorporating previously related edits~\citep{N36}, self-debugging algorithms~\citep{N51}, among others. The third type applies LLMs to vulnerability mining or other related tasks~\citep{N18, N25, N93, N94}, including decompilation work~\citep{N127}, security analysis of hardware designs~\citep{N57}, and black-box testing~\citep{N63}.

\subsection{Code Evaluation}

\noindent \textbf{Definition:} Code evaluation is a crucial task in software engineering that help ensure code quality, reliability, and expected functionality. Code evaluation aims to perform static analysis on the code to identify potential issues and improvement points. When using LLM for code assessment, LLM analyzes the nature of the given code based on the requirements, such as detecting code plagiarism or assessing code compliance with coding standards. Therefore, tasks like code cloning and code smell detection also fall under this category within the scope of this article's definition. 

Code testing involves executing the code and checking if its behavior aligns with expectations to validate its correctness. However, code testing can be time-consuming and labour-intensive. Exploring the application of LLMs to automatically generate effective test cases based on a given code or directly evaluate the quality of a given code has attracted significant research attention.

In current research on applying LLMs to code testing, a significant portion of the work focuses on test case generation. For example, empirical studies have explored the unit test generation capabilities of LLMs like ChatGPT~\citep{N40, N130} or utilizing LLMs to generate test cases~\citep{N131, N91, N128, N129, N44}.

Additionally, several works have proposed code testing frameworks utilizing LLMs based on different usage scenarios~\citep{N62, N43}, such as combining LLMs with fuzz testing~\citep{N135} or black-box testing~\citep{N63}. Furthermore, a few works have developed LLM-based testing assistants or code testing models~\citep{N86, N90}.

\subsection{Code Management}

\noindent \textbf{Definition:} Code management is a crucial aspect of the software engineering development and maintenance process. When using LLM for code management, LLM analyzes the given code based on the requirements, such as version control~\citep{N163}, collaborative management~\citep{N164}, and release management of source code during software development~\citep{N162}.

Code management is complex and challenging. Parallel development and collaboration can lead to code conflicts. Moreover, in large-scale projects, branch management and version iterations significantly increase the difficulty of code management. Traditional code management tools, such as Git~\citep{N160}, provide a powerful set of commands and rules to handle version control and branch management tasks. However, using these tools still requires developers to have expertise, especially when dealing with complex merge conflicts and branch management. 

Due to the various challenges in code management tasks, some cutting-edge works aim to leverage the power of LLMs to alleviate the complexity of the code management process, and even achieve automated code management without manual intervention~\citep{N46, N47}. For example, Toufique et al.\citep{N92} evaluates the effectiveness of LLMs in assisting engineers in managing security incidents in cloud services, while Shrivastava et al.\citep{N19} addresses the issue of LLM models struggling to understand the context present in repositories. Additionally, some works based on LLMs have developed code management tools or frameworks to assist code managers, aiding in version management~\citep{N50} and personnel training~\citep{N81}.

\subsection{Q\&A Interaction}

\noindent \textbf{Definition:} Interaction between humans and tools has always been a focus in the field of software engineering, as good interaction can enhance task performance~\citep{N167}.The wide-ranging application and research of LLM have led to the emergence of this task as a relatively independent research area. In the scope defined in this article, prompt engineering is also considered part of this task.

Before the widespread application of LLMs, an important way for developers to obtain information and solve problems was through Q\&A website, e.g., Stack Overflow\footnote{www.stackoverflow.com/}~\citep{N168}. The emergence of LLMs changed this by being able to answer users' questions, including professional knowledge in software engineering. As a promising new tool to help developers solve code issues, LLMs also gave rise to much research on how to improve the efficiency and convenience of Q\&A Interaction~\citep{N166}. Furthermore, since the output generated by LLMs is influenced by the structure and content of user-provided prompts, research on prompts, known as prompt engineering~\citep{N7}.

It is important to note that this section focuses on investigations related to Q\&A Interaction and prompt engineering in the context of software engineering.

This body of work can also be categorized into two main types. The first type focuses on the interaction between software practitioners (developers, beginners, etc.) and LLMs, and involves the development of prototype systems or interaction frameworks~\citep{N6, N2, N48}. Among them, Zamfirescu-Pereira et al.~\citep{N2} discusses the role of non-AI practitioners in ``user cue engineering" and designs BotDesigner, a cue-based chatbot design tool; Ross et al.~\citep{N6} demonstrates the role and potential of developer-LLM interactions for processes such as software development, through interviews with 42 software developers; and Cai et al.~\citep{N48} describes Low-code LLM, a framework for human-LLM interactions, to better support visual programming.

The second type consists of research-oriented work, which can be further divided into several directions. The first direction evaluates the interaction between LLMs and software developers~\citep{N138}, such as whether LLMs address the same parts of natural language descriptions as developers~\citep{N26}, or whether they can act as a DevBot~\citep{N10}.

The second direction primarily focuses on prompt engineering~\citep{N5, N7, N69}, aiming to design more efficient prompt formats or automatically populate prompt content based on different subtasks and objectives. The third direction addresses security and efficiency issues in LLM interaction with developers~\citep{N65, N66}.

\subsection{Other Works}

In addition to the aforementioned topics, there are other works that combine LLMs with software engineering. These works either discuss the performance of LLMs in specific subtasks~\citep{N13,N27,N8}, such as visualization~\citep{N56}, information extraction~\citep{N45,N39}, and modeling~\citep{N136}, propose their own solutions to existing problems, such as addressing performance issues~\citep{N30}, develop tools or datasets, such as code-text datasets~\citep{N38, N82}, or identify issues related to LLMs~\citep{N12, N67}. Additionally, some works focus on exploring the potential and applications of LLMs in the field of education~\citep{N59}.

%% file: RQ2.tex
\newcommand{\performance}{Author's Confidence}
\newcommand{\performancelow}{author's confidence}
\newcommand{\well}{High Confidence}
\newcommand{\limit}{Low Confidence}

\begin{table}[htbp]
\caption{Part 1 - The \performancelow\  of LLMs in software engineering tasks.}
\setlength{\tabcolsep}{2mm}
\renewcommand\arraystretch{1.2}
\label{tab:RQ2-1}
\begin{tabular}{p{40pt}  p{40pt} p{70pt} p{70pt}   p{100pt}}
\toprule
\textbf{Paper}         & \textbf{Source}                  & \textbf{LLM(s)}                                              & \textbf{Subject of evaluation}                                                                                                                       & \textbf{\performance\ }                                                                                                                                                     \\ \midrule
\citep{N11}  &preprint  & ChatGPT                                        &  code generation, program repair,  and code summarization                                                                                          & \uwave{\well}\                                                                                                                      \\  \midrule
\citep{N1}  &preprint    & ChatGPT                                       & 15 common software engineering tasks                                                                                                             &  Well done: log summary, anaphora parsing, code summarization (method name generation) code clone detection , etc  Not doing well: code vulnerability detection, etc     \\ \midrule
\citep{N17}  &preprint  & ChatGPT                                         &  syntax understanding, static behaviour understanding, dynamic behaviour understanding, capacity to comprehend code syntax,  and capacity to comprehend semantic structures   & Well done: Code syntax understanding Not doing well: Code semantic understanding                                                                             \\ \midrule
\citep{N22}  &preprint      & Codex, and GPT-3.5                              & code interpretation                                                                                                                                  & \well\                                                                                                                                                 \\ \midrule
\citep{N44}  &preprint  & ChatGPT                                        & code defect detection                                                                                                                            & \limit                                                                                                                                             \\ \midrule
\citep{N70}  &preprint   & Codex                              & software reverse engineering                                                                                                                     & \limit                                                                                                                                             \\ \midrule
\citep{N83} &preprint  & ChatGPT                                         & test case generation                                                                                                                                  & \well\                                                                                                                                                 \\ \midrule
\citep{N71} &preprint    & Codex                                         & code defect detection                                                                                                                            & \limit                                                                                                                                             \\ \midrule
\citep{N95}   &preprint    & Codex                                        & code generation                                                                                                                                  & \well\                                                                                                                                                 \\ \midrule
\citep{N130} &preprint   & ChatGPT                                         & test case generation                                                                                                                             & \underline{\limit}                                                                                                                   \\ \midrule
\citep{N74}  &ICER  & Codex                                           & code interpretation                                                                                                                                & \well\                                                                                                                                                 \\ \midrule
\citep{N125} &ICSTW   & ChatGPT                                          & software testing                                                                                                                                 & \well\                                                                                                                                                 \\ \midrule
\citep{N139}  &preprint & GPT-4                                             & code generation                                                                                                                                  & \well\  and obvious progress                                                                                                                           \\ \midrule

\end{tabular}
\end{table}

\begin{table}[htbp]
\caption{Part 2 - The \performancelow\  of LLMs in software engineering tasks.}
\setlength{\tabcolsep}{2mm}
\renewcommand\arraystretch{1.2}
\label{tab:RQ2-2}
\begin{tabular}{p{40pt}  p{40pt} p{70pt} p{70pt}   p{100pt}}
\toprule
\textbf{Paper}               & \textbf{Source}            & \textbf{LLM(s)}                                              & \textbf{Subject of evaluation}                                                                                                                       & \textbf{\performance\ }                                                                                                                                                     \\ \midrule
\citep{N79}  &preprint & Codex                                            & code interpretation                                                                                                                              & \limit (robustness)                                                                                                                                 \\ \midrule
\citep{N84}    &preprint              & GPT-4                                            & code analysis                                                                                                                                    & \underline{\limit}                                                                                                                     \\ \midrule
\citep{N93} &preprint & Codex                                            & program repair                                                                                                                         & \well\                                                                                                                                                 \\ \midrule
\citep{N94} &preprint  & GPT-Neo-125M/1.3B/2.7B, GPT-J-6.7B, GPT-NeoX-20B, Codex-12B, CodeT5-220M and INCODER-1.3B/6.7B & program repair                                                                                                                         & \well\                                                                                                                                                 \\ \midrule
\citep{N115} &preprint & Codex-12B, CodeGen-16B, InstructGPT, and ChatGPT      & code generation                                                                                                                                  & \underline{\limit} (robustness)                                                                                                         \\ \midrule
\citep{N116} &ISAC & ChatGPT                                          & code generation                                                                                                                                  & Ambiguity                                                                                                                                                    \\ \midrule
\citep{N113} &preprint & code-davinci-002                                 & code generation                                                                                                                                  & \underline{\limit}                                                                                                                     \\ \midrule
\citep{N121} &DATE & CodeGen -2B/6B/16B, and Codex                                 & code generation                                                                                                                                  & \underline{\limit}                                                                                                                     \\ \midrule
\citep{N126} &preprint & ChatGPT                                          & code generation                                                                                                                                  & \uwave{\well}\                                                                                                                         \\ \midrule
\citep{N134} &IWAPR & Codex                                            & program repair                                                                                                                                  & \well\                                                                                                                                                 \\ \midrule
\citep{N76} &CHI & Codex                                            & program repair                                                                                                                                   & \underline{\limit}                                                                                                                     \\ \midrule
\citep{N132} &preprint & ChatGPT                                          & Text-to-SQL                                                                                                                                      & \uwave{\well}\                                                                                                                         \\ \midrule
\citep{N133} &preprint & Codex                                            & Text-to-SQL                                                                                                                                      & \well\                                                                                                                                                 \\ 
\midrule
\citep{N176} &preprint & GPT-4                                            & Vulnerable detect                                                                                                                                      & \well\                                                                                                                                                 \\   \hline
\end{tabular}
\end{table}

\begin{table}[h]
\caption{The datasets and evaluation metric in papers.}
\setlength{\tabcolsep}{2mm}
\renewcommand\arraystretch{1.2}
\label{tab:RQ2-3}
\scalebox{0.9}{
\begin{tabular}{lll}
\hline \textbf{Paper} & \textbf{DataSet} & \textbf{Evaluation metric} \\
\hline \citep{N11} & Leetcode+Refactory & Correct Rate  \\
 \citep{N1} & Multi-dataset & Successful Rate \\
 \citep{N17} & A new dateset & Effective \\
 \citep{N22} & A new dateset & Successful Rate \\
 \citep{N44} & QuixBugs & Correct Rate \\
 \citep{N70} & Program Source Templates & Correct Rate \\
 \citep{N83} & An example study & Effective \\
 \citep{N71} & CWE-787\&89+ExtractFix & Effective \\
 \citep{N95} & MeMo+ A new dataset & Correct Rate \\
 \citep{N130} & Defects4J+A new dataset & Correct Rate, Robustness \\
 \citep{N74} & A new dateset & Sensible, Novel, Solution \\
 \citep{N125} & A new dateset & Correct Rate \\
 \citep{N139} & MCQ & Correct Rate \\
 \citep{N79} & AdvGLUE+GeoQuery+Scholar & Robustness \\
 \citep{N84} & An example study & Correct Rate \\
 \citep{N93} & LMdefects & Effective, Correct Rate, Robustness \\
 \citep{N94} & Defects4J+QuixBugs+ManyBugs & Correct Rate \\
 \citep{N115} & AOJ & Solve Rate \\
 \citep{N116} & A new dateset & Quality \\
 \citep{N113} & A new benchmark Suite & Effective(Score) \\
 \citep{N121} & A new dateset & Quality \\
 \citep{N126} & A new dateset & Security \\
 \citep{N134} & QuixBugs & Correct Rate \\
 \citep{N76} & CWE-787\&89+HumanEval & Effective \\
 \citep{N132} & 9 dataset & Effective \\
 \citep{N133} & GeoQuery+Scholar & Correct Rate \\
 \citep{N176} & A new dateset & Effective(Score) \\
\hline
\end{tabular}
}
\end{table}

\section{Performance of LLM in SE Tasks}
\label{sec:LLM Better}

In this section, we primarily discuss \textbf{RQ2}. First, we screened papers from our collection that evaluated the performance of LLMs on software engineering tasks. Next, we extracted the LLMs used and software engineering tasks targeted in these works. Finally, some works in Section~\ref{sec:LLM Meet SE} also evaluated and discussed the performance of LLMs on some specific tasks. Therefore, we will summarize these works here and emphasize their evaluation results.

A significant portion of the work conducted has empirically analyzed the performance of ChatGPT, one of the most popular LLM models, as a programming assistant~\citep{N11, N1, N44, N132}. These studies have found that ChatGPT's performance varies across different tasks. For instance, it performs well in tasks such as log summarization, referential resolution, and code summarization, but struggles in vulnerability detection and test case generation. Particularly in vulnerability detection, ChatGPT finds it challenging to identify subtle code differences when two versions have similar syntax~\citep{N44}. In some tasks such as Text-to-SQL~\citep{N132}, answering software testing questions~\citep{N125}, and test case generation~\citep{N130}, although ChatGPT did not achieve outstanding performance, the authors still maintain a positive outlook. Some studies also highlight the limitations of ChatGPT's attention scope~\citep{N1}.

Furthermore, some works analyze ChatGPT's performance in software engineering tasks from different perspectives. For instance, Ma et al.\citep{N17} investigates ChatGPT's understanding of code syntax and semantic structure, concluding that while ChatGPT excels in understanding code syntax (e.g., Abstract Syntax Trees), it faces difficulties in understanding code semantics, especially dynamic semantics. Feng et al.\citep{N116} explores ChatGPT's code generation abilities through analyzing comments on Twitter and Reddit, examining people's sentiment towards ChatGPT's code generation capabilities.

\begin{table}[h]
\caption{Model and Model structure in papers.}
\setlength{\tabcolsep}{2mm}
\renewcommand\arraystretch{1.2}
\label{tab:RQ2-4}
\begin{tabular}{ll}
\hline \textbf{Model} & \textbf{Model structure} \\
\hline Codex & Decoder-Onlys (Causal Decoder) \\
 ChatGPT/GPT-3.5 & Decoder-Onlys (Causal Decoder) \\
 GPT-4 & Decoder-Onlys (Causal Decoder) \\
 GPT-Neo & Decoder-Onlys (Causal Decoder) \\
 GPT-J & Decoder-Onlys (Causal Decoder) \\
 GPT-NeoX & Decoder-Onlys (Causal Decoder) \\
 CodeT5 & Encoder-decoder \\
INCODER & Encoder-decoder \\
 CodeGen & Decoder-Onlys (Causal Decoder) \\
InstructGPT & Decoder-Onlys (Causal Decoder) \\
 Flan-T5 & Encoder-decoder \\
UnixCoder & Encoder-decoder \\
\hline
\end{tabular}
\end{table}

There are also detailed evaluations of LLMs' performance in specific tasks, such as reverse engineering~\citep{N70}, code explanation~\citep{N79}, code analysis~\citep{N84}, and vulnerability repair~\citep{N71}. These studies generally provide more critical conclusions, suggesting that LLMs still lag behind state-of-the-art methods in these tasks. However, two works evaluating LLMs in automated program repair~\citep{N93, N94} present very positive findings. Additionally, several evaluations on specific tasks yield more positive conclusions or affirm the potential of LLMs in those tasks, such as code generation~\citep{N76, N113, N121} and error fixing~\citep{N134}. \citep{N83} evaluates the ability of large models to generate test cases on a simple game, reporting positive results. \citep{N95} acknowledges the performance of LLMs in code generation capabilities.

Moreover, a considerable portion of the research evaluates the performance of LLMs in education and programming courses~\citep{N52, N22, N74, N139}, with positive feedback. Additionally, compared to GPT-3, GPT-4 has made significant advancements~\citep{N139}.

A small number of works focus on the robustness and security issues of LLMs in solving programming problems~\citep{N115, N126}, and they have yielded important findings as well.

We organize the conclusions derived from the work above, as shown in Table~\ref{tab:RQ2-1} and Table~\ref{tab:RQ2-2}.

We mainly screen the evaluation results from the abstract and conclusion sections of the papers. Due to the inconsistency in evaluation methods and criteria for LLMs across different papers, we cannot directly standardize the assessment of a LLM's performance. We convert the results from the original papers into the "author's confidence" of LLMs' performance in completing SE tasks. 
We categorized the results into: \well , \limit and Ambiguity. 
That is, if authors consider the performance of LLMs to be good, which is "\well"; if authors consider the performance of LLMs to be limited, which is "\limit"; If the authors cannot determine LLM's performance, then it is "Ambiguity".

It is worth noting that the \underline{Author’s Confidence} and the \uwave{Author’s Confidence} respectively indicate that, although the evaluated LLM(s) in the article did not receive a \well\  on the task, the article still provided a positive attitude towards its future potential; and that 
although LLMs is currently performing well, there are still limitations or even not good enough.

Additionally, we have presented the datasets and evaluation metrics used in the aforementioned papers, as shown in Table~\ref{tab:RQ2-3}.

From the table and the above articles, we can reach a preliminary conclusion:
\begin{tcolorbox}[title = {Findings}] 
\begin{itemize}
\item Code generation, being a challenging task, current LLMs often do not achieve a ``ready-to-use" level. The \performancelow\  to LLMs on this task is not enough. However, encouragingly, some articles\citep{N113, N121}, even though the evaluation conclusion did not affirm LLMs, still offered a positive attitude;
\item On tasks like program repair, Text-to-SQL, 
authors usually have high confidence in the performance of LLMs.
\item In program vulnerability detection, 
authors usually have low confidence in the performance of LLMs.
\item Although there is a contradiction in the conclusions of \citep{N83} and \citep{N130} about LLMs on the task of test case generation, they both expressed a relatively positive attitude, still believing that LLMs have potential in this task;
\item In tasks like code summarization and code explanation, LLMs usually perform well, but lack robustness;
\item According to the results of \citep{N139}, newer LLMs have made significant improvements in the task of code generation.
\end{itemize}

\end{tcolorbox}

In summary, we can reach a fundamental conclusion: LLMs perform well and received high confidence from the authors on some software engineering tasks that require understanding of code syntax, such as code summarization and code repair; on some tasks that require understanding of code semantics, such as code generation and vulnerability detection, they typically do not perform well enough; LLMs continue to improve with the iteration of versions/models, and still possess great potential.

\textbf{Therefore, at the current stage, LLMs still cannot fully achieve the level of professional human programmers in handling software engineering tasks, but they can serve as assistants to software developers, such as answering development questions and providing code examples.}

%% file: RelatedWork.tex
\section{Related work}
\label{sec:related work}

\subsection{Other works on reviewing LLM}
The tremendous potential of LLM has attracted numerous investigations regarding its applications, either within the field of LLM itself or in specific domains. 
In this section, we will showcase these research efforts and explain their distinctions from our work. 
Our work focus on systematically investigate, analyze, and compile the research progress of LLM in the context of software engineering tasks.

Zhao et al.~\citep{G1_zhao2023survey} is a detailed article that introduces the background, development history, technical roadmap, and latest advancements of Large Language Models (LLM). The article primarily focuses on large-scale models (with a size greater than 10B) and does not cover early pre-trained language models such as BERT and GPT-2. The research primarily revolves around four main aspects of LLM: pre-training, fine-tuning, applications, and performance evaluation. Additionally, the author considers certain tasks in the software engineering domain as fundamental capabilities of LLM, for instance, treating Code Synthesis as part of the Language Generation capability. Consequently, the article does not delve into a comprehensive discussion and summary of LLM's application, performance, and limitations in software engineering tasks.

Gozalo-Brizuela et al.~\citep{N96}, on the other hand, classifies generative AI models based on their input and output formats, dividing them into nine categories as shown in Table~\ref{tab:96paper}. The paper demonstrates the capabilities of generative AI models through these classifications. While the author believes these models possess significant creativity and potential, he also acknowledges the numerous constraints they face, particularly in terms of data acquisition. Additionally, concerns were raised over the significant computational resource consumption during their training and the time cost of model construction. Despite summarizing the capabilities of some generative AI models, the paper does not focus on LLM, nor does it discuss the performance of LLM in specific tasks, including software engineering tasks.

\begin{table}[h]
\centering
\footnotesize
\caption{A taxonomy of the most popular generative AI models in \citep{N96}}
\setlength{\tabcolsep}{1.5mm}
\renewcommand\arraystretch{1}
\label{tab:96paper}
\begin{tabular}{p{150pt}   p{150pt}}
\toprule
\textbf{Generative AI models categories} & \textbf{Examples}                                     \\ \midrule
Text-to-image Models                     & DALL·E 2, IMAGEN, Muse        \\ \midrule
Text-to-3D models                        & Dreamfusion, Magic3D                          \\ \midrule
Image-to-Text models                     & Flamingo, VisualGPT                           \\ \midrule
Text-to-Video models                     & Phenaki, Soundify                             \\ \midrule
Text-to-Audio models                     & AudioLM, Jukebox, Whisper                      \\ \midrule
Text-to-Text models                      & ChatGPT, LaMDA, PEER \\ \midrule
Text-to-Code models                      & Codex, Alphacode                              \\ \midrule
Text-to-Science models                   & Galactica, Minerva                           \\ \midrule
Other models                             & AlphaTensor, GATO, ChatBCG            \\ \bottomrule
\end{tabular}
\end{table}

Liu et al.~\citep{N108} provides a comprehensive review of ChatGPT and GPT4, highlighting their potential applications and contributions in the field of Natural Language Processing (NLP). Meanwhile, it outlines several potential ethical issues related to the development and use of LLMs, advocating for a focus on addressing these ethical concerns, exploring new applications, and ensuring the responsible use of ChatGPT and GPT-4. Fan et al.~\citep{N98} claims that they conducted a bibliometric analysis of over 5,000 LLM research papers from 2017 to early 2023, investigating their applications across various fields. The paper calls for accelerated cooperation among stakeholders, such as government agencies, universities, companies, infrastructure service providers, etc., to responsibly develop and apply LLMs.

Wei et al.~\citep{N97} provides a comprehensive overview of traditional language models (CLMs) and their successors, pre-trained language models (PLMs). CLMs are designed to predict language sequence probabilities in a causal manner, while PLMs can be fine-tuned for causal sequence modeling and downstream applications. While the article does not delve into the application of large models in the field of software engineering, it offers a brilliant overview of the current state and development progress of large models, and clearly points out future research directions for these models.

Additionally, some research work has reviewed and empirically analyzed the application of LLM in a specific field or particular tasks.

Li et al.~\citep{N101} provides an overview of representative research achievements in text generation based on PLMs, reviews various evaluation metrics, open-source libraries, and common applications, with the aim to assist practitioners in assessing, selecting, and utilizing appropriate PLMs, and proposes some future research directions. Yang et al.~\citep{N109} investigates the application of Transformer-based PLMs for the Controllable Text Generation (CTG) task, summarizing typical applications, key methodologies, and evaluation systems of PLMs in CTG. Min et al.~\citep{N103} explores three trending paradigms of using pre-trained language models for NLP, They are Pre-train then Fine-tune, Prompt-based Learning, and NLP as Text Generation. Simultaneously, the paper mentions that their theoretical understanding of these paradigms is preliminary.

Wang et al.~\citep{N102} primarily summarizes the latest progress of PLMs in the biomedical field and their applications in downstream biomedical tasks, and discusses the development trends and directions. In response to the existing LLM-based recommendation systems, Wu et al.~\citep{N104} proposes a classification that divides these models into two major paradigms, namely, Discriminative LLM for Recommendation (DLLM4Rec) and Generative LLM for Recommendation (GLLM4Rec). The paper systematically reviews and analyzes the existing LLM-based recommendation systems in these two paradigms and discusses their methods and performance.

Other research mainly investigates and analyzes the shortcomings of LLMs in their applications. For instance, Meade et al.~\citep{N100} survey and summarize the bias mitigation techniques in pre-trained language models. The paper empirically studies the five currently popular bias mitigation techniques and proposes three intrinsic bias benchmarks to quantify the effectiveness of each technique. Huang et al.~\citep{N105} comprehensively describes the reasoning capabilities of LLMs, including technologies to improve and induce reasoning capabilities in these models, methods and benchmarks for assessing reasoning capabilities, and suggestions for future directions. As LLMs sometimes provide unrealistic yet seemingly plausible predictions (termed as hallucinations, see~\citep{N106-1}), Mialon et al.~\citep{N106} reviews two methods to enhance the abilities of LLMs, namely, by leveraging reasoning skills and invoking external tools, aiming to improve context and curb hallucinations. Xu et al.~\citep{N107} pays special attention to the inference stage during the construction of LLMs and reviews the current state of model compression and acceleration techniques, including benchmarks, metrics, and methods.

Zan et al.~\citep{N85} investigates the performance of 27 large models in the field of generating code from a natural language description (or NL2Code). The main contribution of the paper is an intuitive comparison of the NL2Code capabilities of LLMs on the HumanEval benchmark. However, its research is limited to the software engineering task of code generation. It does not summarize the applications of other LLMs in code generation, nor does it investigate other software engineering tasks, such as code conversion.

Wong et al.~\citep{N110} introduces some popular Language Model-based Learning (LLM) approaches and their applications in downstream tasks related to AI-assisted programming. The tasks covered in the article include code generation, code completion, code translation, code refinement, code summarization, defect detection, and clone detection. However, it is worth noting that the AI-assisted methods discussed in this article are not limited to LLM but also encompass various other AI techniques. Furthermore, the focus of the article is solely on AI-assisted programming tasks.

\citep{Nn23} provides a systematic literature review of the intersection between SE and deep learning (DL). It includes 128 references covering diverse SE domains and tasks. The article presents a detailed research roadmap, depicting the current state and application of DL techniques in SE tasks. It also analyzes and discusses future directions for the intersection of DL and SE research. The motivation and classification of SE tasks provided in the paper offer valuable insights for our article. The distinction lies in that the reviewed paper focuses on DL, while our article specifically reviews LLMs.

\subsection{Pre-training model for Software Engineering Tasks}

Previous works have applied pre-trained language models (PLM) to software engineering tasks, providing a solid foundation and guidance for the current application of PLMs in SE tasks. For instance, ~\citep{Nn10} used a pre-trained language model to generate variable names. ~\citep{Nn19} utilized a PLM for bug classification. ~\citep{Nn21} introduced DietCode, a lightweight approach that leverages large pre-trained models for source code generation. ~\citep{Nn22} employed a PLM for automatic comment generation. Additionally, there have been works focusing on prompt learning, such as ~\citep{Nn16}.

There are also several works that have evaluated the performance of PLMs in software engineering tasks. ~\citep{Nn3} studied the capabilities of pre-trained models in code comprehension and investigated the robustness of pre-trained models by examining their performance under adversarial attacks. ~\citep{Nn4} explored the feasibility of using pre-trained models for automatic repair of merge conflicts (both textual and semantic). ~\citep{Nn5} conducted an empirical analysis of the performance of the T5 model in code-related tasks. ~\citep{Nn6} compared and evaluated the accuracy and efficiency of three BERT architectures in linking issues and commits for bug triaging in open-source projects. ~\citep{Nn18} evaluated whether pre-trained language models encode the entire grammar structure of programming languages. ~\citep{Nn20} employed probes to identify if models lack certain code properties (comprehension). ~\citep{Nn8} performed source code analysis of large-scale code models to showcase the interpretability features of Code LLMs. ~\citep{Nn9} validated the ability of the T5 model in automating code review tasks and provided positive conclusions, demonstrating its superiority over previous DL models.

Due to the relatively smaller parameter count of PLMs compared to LLMs, they may lack the powerful code generation capabilities, leading to potential issues in the generated code. To address this, some works have combined PLMs with program analysis techniques to enhance their performance in SE tasks. ~\citep{Nn7} employed post-processing techniques based on program analysis to improve the reliability of LLM code generation. ~\citep{Nn11} proposed a bridging approach between pre-trained models and code-related tasks to enhance the capabilities of pre-trained models in tasks such as code comprehension. ~\citep{Nn12} introduced SPT-Code, a sequence-to-sequence pre-trained model specifically designed for source code. ~\citep{Nn13} attacked LLMs using adversarial input transformations to guide the models to produce incorrect outputs. ~\citep{Nn14} presented AutoSC for code completion tasks. ~\citep{Nn15} introduced CugLM, and ~\citep{Nn17} proposed a model compression technique to facilitate better utilization of Code LLMs by developers.

%% file: Conclusion.tex
\section{Threats to validity}
\subsection{Internal Validity}

Sections 2 and Sections 4 rely on manual screening of papers and analysis of paper content, respectively. As a result, there may be internal subjectivity in the results of these screenings and analyses. To mitigate subjective bias and enhance the reliability of our results, we used a multi-person protocol in both sections, where multiple participants performed separate analyses and aligned the results to ensure that any disagreements were adequately resolved. Notably, the participants in these tasks (including non-co-authors) all had more than 3 years of software engineering research experience.

In Section 3, Our classification method of the SE tasks serves as an early-stage exploration to better structure this emerging field. We acknowledge that our classification may not fully encompass all aspects of software engineering tasks, and we agree that such a classification should be grounded in practice and mapped to actual research papers, rather than derived from conference categories alone. However, we believe our classification still captures a significant portion of the most important tasks in software engineering.

In Section 5, we focus on the performance of LLM on SE tasks. However, there may be differences in different works for LLM evaluation perspectives, such as using different datasets and oriented to different SE tasks. So there may be differences in the results in different literatures. To mitigate this effect, we organize the information about the evaluation results, evaluation datasets, and evaluation perspectives of different works to help readers understand the variability of different works.

\subsection{External Validity}

Currently, LLM technology is evolving rapidly, as is LLM research in the field of software engineering. Although our findings cannot encompass future works, our work is still valuable in revealing the current direction of LLM on SE tasks. Meanwhile, our categorization of LLM in SE tasks remains applicable within the current scope of research.

The second external threat pertains to subtle boundaries that exist between software engineering tasks, such as code cloning and code smells. These tasks may be relevant to multiple software engineering domains, which introduces potential ambiguity in their classification. To mitigate this threat, we have provided explicit definitions for each software engineering task and clarified the scope covered by the task classifications. For example, we have defined code cloning and code smell tasks within the domain of code assessment.

Another external threat is the inherent limitations of the literature review process itself. Due to the impracticality of verifying the accuracy of every work reviewed in the literature, there may be some bias in the results presented for RQ2. To mitigate the impact of this threat, we have provided detailed information regarding the evaluation details of each work surveyed for RQ2. This includes information such as the dataset used, the emphasis of the evaluation, the target of evaluation, and the evaluation results. We believe that if LLM consistently performs poorly in almost all evaluation studies for a specific task, it can be tentatively concluded that LLM performs poorly in that task. Conversely, if all evaluation studies provide positive assessments for a specific SE task, we can conclude that LLM performs well in that task. We have provided a fair summary of the evaluation of LLM across different SE tasks. Furthermore, we address any contradictory evaluations found in the current body of evaluation studies in our conclusion.

\section{Conclusion and Future Work}
\label{sec:conclusion}

This paper comprehensively reviews the applications of Large Language Models (LLMs) in software engineering. Firstly, we collected and screened 123 works and literature related to the intersection of software engineering and LLM. Secondly, by categorizing the selected literature based on software engineering tasks, we revealed the research focus and identified existing deficiencies in the integration of LLM with various software engineering tasks. Finally, we carefully examined and summarized papers evaluating the performance of LLMs in software engineering tasks, exposing their capabilities and limitations and providing directions for future research and optimization.

Current works also reveal some future directions worth discussing:  (1) We can see that a large part of the work in Section~\ref{sec:LLM Meet SE} proposes methods to improve the performance of LLMs on one or several software engineering tasks. Although most of them do not provide detailed evaluations or discussions on the performance of LLMs on these tasks, this might suggest that the current performance of LLMs on these tasks is not good enough or not stable; (2) Most current evaluations are based on general large models, such as ChatGPT, and detailed evaluations of code-centric large models like Codex are still lacking; (3) Do we need to fine-tune large models for specific software engineering tasks to create large model products tailored for specific tasks? We will gradually seek the answers to these questions in the future.

\section{Conflict of Interests}

The authors declared that they have no conflict of interest exits in the submission of this manuscript, and manuscript is approved by all authors for publication. I would like to declare on behalf of my co-authors that the work described was original research that has not been published previously and is not under consideration for publication elsewhere. All the authors listed have approved the manuscript.

We declare that we do not have any commercial or associative interest that represents a conflict of interest in connection with the work submitted.

\section{Data availability statements}

The list of the literature surveyed in this paper is open source and is available in \url{https://github.com/KevinHeiwa/LLM-SE-Paper-List}.

\begin{acknowledgements}
This work was supported by the National Natural Science Foundation of China under Grant 62306344, and Guangdong Basic and Applied Basic Research Foundation under Grant 2024A1515010253.
\end{acknowledgements}